\documentclass[graybox]{svmult}

\usepackage{mathptmx}       
\usepackage{helvet}         
\usepackage{courier}        
\usepackage{type1cm}        
\usepackage{makeidx}         
\usepackage{graphicx}        
\usepackage{multicol}        
\usepackage[bottom]{footmisc}

\makeindex             

\begin{document}

\title*{Lovelock theory, black holes and holography}
\author{Jos\'e D. Edelstein}
\institute{Jos\'e D. Edelstein \at Department of Particle Physics and IGFAE, University of Santiago de Compostela, E-15782\\ Santiago de Compostela, Spain\\ Centro de Estudios Cient\'{\i}ficos CECs, Av. Arturo Prat 514, Valdivia, Chile\\ \email{jose.edelstein@usc.es}}

\maketitle

\abstract*{Lovelock theory is the natural extension of general relativity to higher dimensions. It can be also thought of as a toy model for ghost-free higher curvature gravity. It admits a family of AdS vacua, most (but not all) of them supporting black holes that display interesting features. This provides an appealing arena to explore different holographic aspects in the context of the AdS/CFT correspondence.}


\abstract{Lovelock theory is the natural extension of general relativity to higher dimensions. It can be also thought of as a toy model for ghost-free higher curvature gravity. It admits a family of AdS vacua, most (but not all) of them supporting black holes that display interesting features. This provides an appealing arena to explore different holographic aspects in the context of the AdS/CFT correspondence.}

\section{Lovelock theory}

While classical gravity seems well-described by the Einstein-Hilbert action, quantum corrections generically involve higher curvature terms. This is the case, for instance, of $\alpha^\prime$ corrections in string theory. On general grounds, higher curvature terms arise in Wilsonian low-energy effective descriptions of gravity.

The inclusion of higher curvature corrections customarily leads to higher order equations of motion. They are consequently argued to be plagued of ghosts. Despite that, David Lovelock tackled the problem some four decades ago finding the most general situation leading to second order Euler-Lagrange equations \cite{Lovelock}. He showed that, whereas in four dimensions General Relativity is the natural answer, higher dimensional scenarios lead to the appearance of higher curvature contributions to the action, on equal footing with the Einstein-Hilbert term. The action of Lovelock theory is given, in $d$ space-time dimensions, by a sum of $K \leq [\frac{d-1}{2}]$ terms,
\begin{equation}
\mathcal{I} = \sum_{k=0}^{K} {\frac{c_k}{d-2k}} \,\mathcal{I}_k ~,
\label{gensum}
\end{equation}
which admit a compact expression in terms of differential forms
\begin{equation}
\mathcal{I}_k = \mathop\int \epsilon_{a_1 \cdots a_{d}}\; R^{a_1 a_2} \wedge \cdots \wedge R^{a_{2k-1} a_{2k}} \wedge e^{a_{2k+1}} \wedge \cdots \wedge e^{a_d} ~,
\label{Ik}
\end{equation}
where $\epsilon_{a_1 \cdots a_{d}}$ is the anti-symmetric symbol, $R^{ab} := d\omega^{ab} + \omega_{\;~c}^a\wedge \omega^{cb}$ is the Riemann curvature $2$-form, computed from the spin connection $1$-form $\omega^{ab}$, and $e^{a}$ is the vierbein $1$-form. By construction, Lovelock theories are intrinsically higher dimensional.

It is easy to see that the first two terms (most general up to $d = 4$) are quite familiar; $\mathcal{I}_0$ gives the cosmological term while $\mathcal{I}_1$ is nothing but the Einstein-Hilbert (EH) action. Their normalization is fixed along this talk as
\begin{equation}
L^2 c_0 = c_1 = 1 ~,
\label{c0yc1}
\end{equation}
or, in terms of the more familiar dimensionfull quantities of General Relativity,
\begin{equation}
\Lambda = - \frac{(d-1) (d-2)}{2 L^2} ~, \quad 16\pi (d-3)!\,G_N = 1 ~,
\label{LambdayG}
\end{equation}
$G_N$ being the Newton constant. For $d \geq 5$, for instance, we have the Lanczos-Gauss-Bonnet (LGB) term \cite{Lanczos} ($c_2 = \lambda\, L^2$),
\begin{equation}
{\mathcal{I}_2} \simeq d^{d}x \, \sqrt{-g} \left( R^2-4R_{\mu\nu} R^{\mu\nu} + R_{\mu\nu\rho\sigma} R^{\mu\nu\rho\sigma} \right) ~,
\label{LGBaction}
\end{equation}
while for $d \geq 7$, we introduce the cubic Lovelock Lagrangian ($c_3 = \mu\, L^4$),
\begin{eqnarray}
{\mathcal{I}_3} & \simeq & d^{d}x \, \sqrt{-g} \left( R^3 + 3 R R^{\mu\nu\alpha\beta} R_{\alpha\beta\mu\nu} - 12 R R^{\mu\nu} R_{\mu\nu} \right. \nonumber \\ [0.5em]
& & \left. +\, 24 R^{\mu\nu\alpha\beta} R_{\alpha\mu} R_{\beta\nu} + 16 R^{\mu\nu} R_{\nu\alpha} R_{\mu}^{~\alpha}  + 24 R^{\mu\nu\alpha\beta} R_{\alpha\beta\nu\rho} R_{\mu}^{~\rho} \right. \nonumber \\ [0.5em]
& & \left. +\, 8 R_{~~\alpha\rho}^{\mu\nu} R_{~~\nu\sigma}^{\alpha\beta} R_{~~\mu\beta}^{\rho\sigma} + 2 R_{\alpha\beta\rho\sigma} R^{\mu\nu\alpha\beta} R_{~~\mu\nu}^{\rho\sigma} \right) ~.
\label{LLcubic}
\end{eqnarray}
By simple comparison of (\ref{Ik}) and, say, (\ref{LLcubic}), the advantages of the so-called first order formalism become manifest. The equations of motion are obtained by varying independently with respect to the vierbein and the spin connection. The latter can be solved by simply setting the torsion $T^a := d e^a + \omega^a_{~b} \wedge e^b$ to zero. This is not the most general solution, but the one we will consider along this talk, since it allows us to make contact with the second order metric formulation of gravity.

The equations of motion, when varying the vierbein, can be cast into the form
\begin{equation}
\epsilon_{a a_1 \cdots a_{d-1}}\; \mathcal{F}_{(1)}^{a_1 a_2} \wedge \cdots \wedge \mathcal{F}_{(K)}^{a_{2K-1} a_{2K}} \wedge e^{a_{2K+1}} \wedge \ldots \wedge e^{a_{d-1}} = 0Ê~,
\label{EOM}
\end{equation}
which neatly displays the fact that these theories admit (up to) $K$ constant curvature maximally symmetric {\it vacua},
\begin{equation}
\mathcal{F}_{(i)}^{a b} := R^{a b} - \Lambda_i\, e^{a} \wedge e^{b} = 0 ~.
\label{Kvacua}
\end{equation}
The {\it effective} cosmological constants turn out to be the (real) roots of the characteristic polynomial $\Upsilon[\Lambda]$,
\begin{equation}
\Upsilon[\Lambda] := \sum_{k=0}^{K} c_k\, \Lambda^k = c_K \prod_{i=1}^{K} \left( \Lambda - \Lambda_i \right) ~.
\label{Upsilon}
\end{equation}
We will see that many important features of these theories and their black hole solutions are governed by this polynomial. Degeneracies arise when its discriminant, $\Delta := \prod_{i<j} (\Lambda_i - \Lambda_j)^2$, vanishes. This is typically associated with symmetry enhancement and/or the emergence of non-generic features of Lovelock theory. Even though they can be fairly interesting (see \cite{Giribet} for a recent example), we will mostly deal with the $\Delta \neq 0$ case throughout this presentation.

For the sake of clarity, let us briefly consider the $K = 2$ case. This amounts to the inclusion of the LGB term which, for instance, arises in superstring theory \cite{BachasPG,KatsP,BuchelMS}. Being quadratic, the roots of the polynomial $\Upsilon[\Lambda]$ can be explicitly sorted out:
\begin{equation}
\Lambda_\pm = - \frac{1 \pm \sqrt{1 - 4 \lambda}}{2 \lambda\, L^2} \qquad {\rm then} \qquad \Delta = 0 \quad\Leftrightarrow\quad \lambda = \lambda_{\rm CS} := \frac{1}{4} ~.
\label{LambdaLGB}
\end{equation}
The CS subscript in $\lambda_{\rm CS}$ amounts for Chern-Simons, since that is the critical value of $\lambda$ for which the theory acquires an extra symmetry (in $d=5$) becoming a gauge theory for the AdS group \cite{Chamseddine}. For $0 < \lambda < \lambda_{\rm CS}$ the theory has two AdS vacua;\footnote{If $\lambda < 0$, there is no a priori lower bound for it and it is clear that $\Lambda_+$ becomes positive.} $\Lambda_+$ is known to be unstable \cite{BoulwareDeser}. For $\lambda > \lambda_{\rm CS}$ there is no AdS vacuum.

The branch corresponding to $\Lambda_-$ is called the EH-branch, since it is continuously connected to the solution of General Relativity when $\lambda \to 0$. It has $\Upsilon'[\Lambda_-] > 0$, which amounts to a positive effective Newton constant. In fact, each vacuum $\Lambda_i$ has a different {\it effective} Newton constant, $G^i_N \sim 1/\Upsilon'[\Lambda_i]$, whose sign coincides with that of $\Upsilon'[\Lambda_i]$. Thus, a given root of Lovelock gravity, $\Lambda_\star$, must satisfy
\begin{equation}
\Upsilon'[\Lambda_\star] > 0 ~,
\label{Upsilonprime}
\end{equation}
in order to correspond to a vacuum that hosts gravitons propagating with the right sign of the kinetic term. Else, if $\Upsilon'[\Lambda_\star]$ is negative, we say that the corresponding vacuum is affected by Boulware-Deser (BD) instabilities.

On top of the maximally symmetric vacua, we are interested in shockwave backgrounds of Lovelock theory. We want to show that in the presence of a shockwave there is room for causality violation \cite{Hofman}. This will raise the question whether all possible values of Lovelock's couplings, $c_k$, lead to physically sensible theories of gravity. The shockwave solution on AdS, with cosmological constant $\Lambda_\star$, reads \cite{Hofman}
\begin{equation}
ds^2_{{\rm AdS},sw} = \frac{L_\star^2}{z^2} \left( -du\,dv + d\mathbf x^2 + dz^2 \right) + F(u)\,\varpi(\mathbf x,z)\,du^2 ~,
\end{equation}
where $z = L_\star^2/r$ is the Poincar\'e radial direction, $u,v=x^0\pm x^{d-1}$ are light-cone coordinates, and $\mathbf x$ are the remaining $d-3$ spatial directions. $L_\star$ is the AdS radius corresponding to the branch on top of which we construct the shockwave, $L_\star \sim (-\Lambda_\star)^{-1/2}$. We should think of $F(u)$ as a distribution with support in $u = 0$, which we will finally identify as a Dirac delta function, $F(u) = \delta(u)$.

The shock wave is parameterized by the function $\varpi(\mathbf x,z)$, obeying
\begin{equation}
2 (d-3) \varpi + (d-6) z \partial_z \varpi - z^2 (\partial_z^2 + \nabla_\bot^2) \varpi = 0 ~,
\end{equation}
where $\nabla_\bot^2$ is the Laplacian in the $\mathbf x$-space. This equation admits the following solutions, whose holographic counterpart will be briefly addressed later. The simplest profile
\begin{equation}
\varpi = \varpi_0\;z^{d-3} ~,
\label{sw1}
\end{equation}
on the one hand, and the $\mathbf x$-dependent solution
\begin{equation}
\varpi = \varpi_0\;\frac{z^{d-3}}{\left(z^2 + (\mathbf x - \mathbf x_0)^2 \right)^{d-2}} ~, \qquad \mathbf x_0 = \frac{\mathbf n}{1 + |\mathbf n|^{d-2}} ~,
\label{sw2}
\end{equation}
where $\mathbf n$ is a unit vector. We will use these shockwave profiles below.

\section{Black holes}

The black holes of Lovelock theory were exhaustively studied in \cite{CamanhoEdelstein3}. In this talk we will just discuss some salient features that are instrumental to their holographic applications. The solutions can be obtained from the ansatz \cite{BoulwareDeser,Wheeler,Wheeler2}
\begin{equation}
ds^2 = - f(r)\, dt^2 + \frac{dr^2}{f(r)} + \frac{r^2}{L^2}\, d \Sigma_{\sigma,d-2}^2 ~,
\end{equation}
where $d\Sigma_{\sigma,d-2}$ is the metric of a $(d-2)$-dimensional manifold, $\mathcal M$, of negative, zero or positive constant curvature ($\sigma =-1, 0, 1$ parametrizing the different horizon topologies). A natural frame is given by
\begin{equation}
e^0 = \sqrt{f(r)}\, dt ~, \qquad e^1 = \frac{1}{\sqrt{f(r)}}\, dr ~, \qquad e^a = \frac{r}{L}\, \tilde e^a ~,
\label{vierbh}
\end{equation}
where $a = 2, \ldots, d-1$, and $\tilde R^{ab} = \sigma\, \tilde e^a \wedge \tilde e^b$. The Riemann 2-form reads
\begin{eqnarray}
& & R^{01} = - \frac12\, f''(r)\; e^0 \wedge e^1 ~, \quad\qquad R^{0a} = - \frac{f'(r)}{2 r}\; e^0 \wedge e^a ~, \nonumber \\ [1em]
& & R^{1a} = - \frac{f'(r)}{2 r}\; e^1 \wedge e^a ~, \qquad\qquad R^{ab} = - \frac{f(r) - \sigma}{r^2}\; e^a \wedge e^b ~.
\label{riemannbh}
\end{eqnarray}
Strikingly enough, if we insert these expressions into the equations of motion, we get after some manipulations a quite simple ordinary differential equation --not for $f(r)$ but for $\Upsilon[g(r)]$,
\begin{equation}
\left[ \frac{d~}{d\log r} + (d-1) \right]\, \left( \sum_{k=0}^{K} c_k\, g^k \right) = 0 ~,
\label{eqgg}
\end{equation}
where $g(r) := \frac{\sigma-f(r)}{r^2}$. It can be straightforwardly solved as
\begin{equation}
\Upsilon[g] = {\rm V}_{\!d-2}\;\frac{M}{r^{d-1}} ~,
\label{LLbhsol}
\end{equation}
where the integration constant, through the Hamiltonian formalism \cite{KastorRT}, can be seen to be the space-time mass $M$ times the volume V$_{\!d-2}$ of the unit constant curvature manifold $\mathcal M$. The black hole solutions are implicitly (and analytically!) given by this polynomial equation. The variation of $r$ translates the $y$-intercept of $\Upsilon[g]$ rigidly, upwards. This leads to $K$ branches, $g_i(r)$, corresponding to the monotonous sections of $\Upsilon[g]$, associated with each $\Lambda_i$: $g_i(r\rightarrow\infty) = \Lambda_i$.

The existence of a black hole horizon requires $g_+ = 0$ for planar black holes, and, since $g_+ = \sigma/r_+^2$,
\begin{equation}
\Upsilon[g_+] = {\rm V}_{\!d-2}\;M\,|g_+|^{(d-1)/2} ~,
\label{Mcurve}
\end{equation}
for spherical or hyperbolic black holes. In the case of non-planar black holes, the curve (\ref{Mcurve}) can intersect the polynomial at different points. Several branches can display black holes with the same mass or temperature. This entails the possibility of a rich phase diagram, provided that the free energy or entropy of these solutions differ, which turns out to be the case \cite{CEGG1,CEGG2,Camanho}.

The plethora of vacua and possibilities for the local behavior of the polynomial $\Upsilon[g]$ lead to a bestiary of black hole solutions that has been analyzed in depth \cite{CamanhoEdelstein3}. We will just review some features of black holes belonging to the EH branch,\footnote{Recall that it is the branch crossing $g=0$ with slope $\Upsilon'[0]=1$.} which are a sort of distorted Schwarzschild-AdS black holes. When real, the effective cosmological constant associated with this branch, $\Lambda_\star$, is negative and so the space-time is asymptotically AdS, regardless of the sign of the cosmological constant appearing in the original Lagrangian.

Even though the EH-branch is just a deformation of the usual Schwarzschild-AdS black hole, it can be a quite dramatic one. For instance, it may happen that the polynomial has a minimum at $g_{\rm min} < 0$, such that $\Upsilon[g_{\rm min}] > 0$. Now, by derivation of (\ref{LLbhsol}) with respect to the radial variable,
\begin{equation}
g' = - (d-1)\,{\rm V}_{\!d-2}\;\frac{M}{r^d}\;\Upsilon'[g]^{-1} ~.
\label{singularity}
\end{equation} 
making clear that the metric is regular everywhere except at $r=0$ and at points where $\Upsilon'[g]=0$. A naked singularity would arise at large radius, $r_{\rm naked}$, where $g(r_{\rm naked}) = g_{\rm min}$. This case was first discussed in \cite{deBoerKP2,CamanhoEdelstein2} for third order Lovelock theory and planar topology, but the same applies in the general case for a vast region of the space of parameters that we call the {\it excluded region} (see Fig.\ref{LL3regions}). We will assume in what follows that the Lovelock couplings do not belong to the excluded region (in the LGB case, this simply means $\lambda \leq 1/4$).

For {\it hyperbolic} or {\it planar} topology, as this branch always crosses $g=0$ with positive slope, it has always a horizon hiding the singularity of the geometry which is located either at $r=0$ [(a) type] or at the value $r_\star$ corresponding to a maximum of $\Upsilon[g]$ [(b) type] for which $g(r_\star) = g_{\rm max} > 0$. Hyperbolic black holes can have a negative mass above a critical value that is nothing but an extremal solution. 

The {\it spherical} case is quite more involved. For high enough mass, the existence of the horizon is ensured, but this is not the case in general. For the (a) type EH-branch the existence of the horizon is certain for arbitrarily low masses if $d>2K+1$. The {\it critical} case, $d=2K+1$, is more subtle. There will be a minimal mass $M_{\rm crit}$ related to the gravitational coupling $c_K$ below which a naked singularity appears \cite{CamanhoEdelstein3}. For high enough orders of the Lovelock polynomial, {\it multi-horizon black holes} can exist but for the critical case, at some point, all of them disappear.

The case of a (b) type branch is simpler. There is a critical value of the mass, $M_\star$, for which the horizon coincides with the singularity, $r_+=r_\star$. Below that mass a naked singularity forms. The simplest example is LGB gravity with $\lambda<0$, where the EH branch has a maximum at $g_{\rm max} > 0$. This is a singularity at finite $r$ that may or may not be naked depending on the value of the mass in relation to $M_\star$,
\begin{equation}
M_\star = \frac{(-2\lambda L^2)^{(d-3)/2}}{2\,{\rm V}_{\!d-2}} (1-4\lambda)  ~.
\end{equation}
For bigger masses we have a well defined horizon while below this bound the singularity is naked.

Some aspects of Lovelock black holes {\it thermodynamics} have been considered in \cite{Cai}. The (outermost) event horizon has a well defined (positive) temperature
\begin{equation}
T = \frac{f'(r_+)}{4\pi} = \frac{r_+}{4\pi}\,\left[(d-1)\,\frac{\Upsilon[g_+]}{\Upsilon'[g_+]}-2\,g_+ \right] ~.
\end{equation}
It is easy to see that large black holes have $M \sim {\rm V}_{\!d-2}\;T^{d-1}$. Then, $dM/dT > 0$ and they can be put in equilibrium with a thermal bath. They are locally thermodynamically stable. In general, this will not happen for small black holes, pointing towards the occurrence of Hawking-Page phase transitions, which have been already studied in the case of LGB gravity \cite{NojiriO,ChoN}. One important feature regarding the classical stability of these black holes is that
\begin{equation}
\frac{dS}{dr_+} = \frac{1}{T} \frac{dM}{dr_+} \simeq r_+^{d-3}\;\Upsilon'\left[g_+\right] ~,
\label{dMdr}
\end{equation}
and, as long as we are in a branch free from BD instabilities, both the radial derivative of the mass and the entropy are positive. This is necessary to discuss classical instability, since the heat capacity reads
\begin{equation}
C=\frac{dM}{dT}=\frac{dM}{dr_+}\frac{dr_+}{dT} ~,
\label{heatcapacity}
\end{equation}
and then the only factor that can be negative leading to an instability is
\begin{equation}
\frac{dT}{dr_+}=-\frac{g_+}{2\pi}\left[(d-2)-\frac{d-1}{2}\frac{\Upsilon[g_+]}{g_+\Upsilon'[g_+]}\left(1+2g_+\frac{\Upsilon''[g_+]}{\Upsilon'[g_+]}\right)\right] ~.
\label{dTdr}
\end{equation}
It is not easy to check classical stability in full generality for non-planar black holes, but in the regimes of {\it high} and {\it low} masses. In the simplest case of planar black holes, the thermodynamic variables do not receive any correction from the higher curvature terms in the action and the expression reduces to the usual formula
\begin{equation}
\frac{dT}{dr_+}=\frac{d-1}{4\pi L^2} ~.
\end{equation}
This expression is manifestly positive. Therefore, these black holes are locally thermodynamically stable for all values of the mass. This is also the case for maximally degenerated Lovelock theories that admit a single (EH-)branch of black holes \cite{CrisostomoTZ}. The entropy can be easily obtained by integrating (\ref{dMdr}),
\begin{equation}
S \simeq r_+^{d-2}\;\left(1+\sum_{k=2}^{K}{k\,c_k\frac{d-2}{d-2k}\, g_+^{k-1}}\right) ~,
\label{entropy}
\end{equation}
and it coincides with the prescription obtained by other means such as the Wald entropy \cite{JacobsonMyers} or the euclideanized on-shell action \cite{MyersSimon}. For planar horizons this formula reproduces the proportionality of the entropy and the area of the event horizon, $S \simeq r_+^{d-2}$, whereas it gets corrections for other topologies. From these quantities we can now compute any other thermodynamic potential such as the Helmholtz free energy, $F=M-TS$, 
\begin{equation}
F \simeq \frac{r_+^{d-1}}{\Upsilon'[g_+]}\sum_{k,m=0}^{K}{\frac{2m-2k+1}{d-2k}\,k\,c_k\,c_m\,g_+^{k+m-1}} ~.
\label{free-energy}
\end{equation}
This magnitude is relevant to analyze the global stability of the solutions for processes at constant temperature. As a function of $g_+$, it has a polynomial of degree $2K-1$ in the numerator. This is the maximal number of zeros that may eventually correspond to Hawking-Page-like phase transitions. Moreover, taking into account that (\ref{free-energy}) is a sum involving the whole set of branches of the theory, phase transitions involving jumps between different branches are expected \cite{CEGG1,CEGG2,Camanho}.

\section{Holography}

The main motivation of our work in Lovelock theory is gaining a better understanding of some aspects of the AdS/CFT correspondence. Since the groundbreaking paper of Juan Maldacena \cite{Maldacena}, evidence has been accumulating towards the validity of the following bold statement: a theory of quantum gravity in AdS space-time is equal to a corresponding (dual) CFT living at the boundary. The relation between both descriptions of the same physical system is holographic.

A key ingredient of this highly nontrivial statement is given by the recipe to compute holographically correlation functions in the CFT \cite{GKP,Witten}. Restricted to the stress-energy tensor, $T_{ab}(\mathbf x)$, it reads
\begin{equation}
\mathcal{Z} \left[ g_{\mu\nu} \right] \approx \exp \left( - \mathcal{I}[g_{\mu\nu}] \right) = \bigg< \exp \bigg( \int\!d\mathbf x ~\eta^{ab}(\mathbf x)\; T_{ab}(\mathbf x) \bigg) \bigg>_{\!\rm CFT} ~,
\end{equation}
where $\mathcal{Z} \left[ g_{\mu\nu} \right]$ is the partition function of quantum gravity, and $g_{\mu\nu} = g_{\mu\nu}(z,\mathbf x)$ such that $g_{ab}(0,\mathbf x) = \eta_{ab}(\mathbf x)$. From this expression, correlators of the stress-energy tensor can be obtained by performing functional derivatives of the gravity action with respect to the boundary metric. This, in turn, is simply given by considering gravitational fluctuations around an asymptotically AdS configuration of the theory.

In the remainder of this presentation, we will investigate the uses of this framework in the case of Lovelock theory and extract some of its consequences.

\subsection{CFT unitarity and $2$-point functions}

Consider a CFT$_{d-1}$. The leading singularity of the $2$-point function is fully characterized by the central charge $C_T$ \cite{OsbornPetkou}
\begin{equation}
\langle T_{ab}(\mathbf x)\, T_{cd}(\mathbf 0)\rangle = \frac{C_T}{\mathbf x^{2(d-1)}}\;\mathcal{I}_{ab,cd}(\mathbf x) ~,
\label{TTcorrelator}
\end{equation}
where
\begin{equation}
\mathcal{I}_{ab,cd}(\mathbf x) = \frac12 \left( I_{ac}(\mathbf x)\, I_{bd}(\mathbf x) + I_{ad}(\mathbf x)\, I_{bc}(\mathbf x) - \frac1{d-1}\, \eta_{ab}\, \eta_{cd} \right) ~,
\end{equation}
whereas $I_{ab}(\mathbf x) = \eta_{ab} - 2\,{x_a\, x_b/\mathbf x^2}$. For instance, $C_T$ is proportional in a CFT$_4$ to the standard central charge $c$ that multiplies the (Weyl)$^2$ term in the trace anomaly, $C_T = 40\, c/\pi^4$.

The holographic computation of $C_T$ was performed in \cite{BuchelEMPSS} for LGB, and in \cite{CEP} for Lovelock theory. According to the AdS/CFT dictionary, it is sufficient\footnote{Other components of the metric fluctuations must be considered as well, but they are irrelevant for our current discussion.} to take a metric fluctuation $h_{xy}(z,\mathbf x) := L_\star^2/z^2 \;\phi(z,\mathbf x)$ about empty AdS with cosmological constant $\Lambda_\star$. Expanding (\ref{gensum}) to quadratic order in $\phi$, and evaluating it on-shell,
\begin{equation}
\mathcal{I}_{\rm quad} = \frac{\Upsilon'[\Lambda_\star]}{2 (- \Lambda_\star)^{d/2}} \int\!d{\bf x} ~z^{2-d} \left(\phi\,\partial_z \phi\right) ~.
\label{Iquad}
\end{equation}
Imposing the boundary conditions $\phi(0,\mathbf x) = \hat\phi(\mathbf x)$, the full bulk solution reads
\begin{equation}
\phi(z,\mathbf x) = \frac{d}{d-2} \frac{\Gamma[d]}{\pi^{\frac{d-1}2} \Gamma\left[\frac{d-1}2\right]} \int\!d{\bf y} ~\frac{z^{d-1}}{(z^2 + |\mathbf x - \mathbf y|^2)^{d-1}} \mathcal{I}_{ab,cd}(\mathbf x - \mathbf y) \;\hat\phi(\mathbf y) ~.
\label{fbulk}
\end{equation}
Plugging this expression into $\mathcal{I}_{\rm quad}$, we obtain
\begin{equation}
\mathcal{I}_{\rm quad} = \frac{C_T}{2} \int\!d{\bf x} \int\!d{\bf y} ~\frac{\hat\phi(\mathbf x)\;\mathcal{I}_{ab,cd}(\mathbf x - \mathbf y)\;\hat\phi(\mathbf y)}{|\mathbf x - \mathbf y|^{2(d-1)}} ~,
\label{Iquadfinal}
\end{equation}
where $C_T$ is the central charge of the dual CFT$_{d-1}$,
\begin{equation}
C_T = \frac{d}{d-2} \frac{\Gamma[d]}{\pi^{\frac{d-1}2} \Gamma\left[\frac{d-1}2\right]} \frac{\Upsilon'[\Lambda_\star]}{(- \Lambda_\star)^{d/2}} ~.
\end{equation}
The upshot of this computation in an AdS vacuum, $\Lambda_\star < 0$, is thought-provoking:
\begin{equation}
C_T > 0 \qquad\Longleftrightarrow\qquad \Upsilon'[\Lambda_\star] > 0 ~.
\end{equation}
The latter inequality, in the gravity side, corresponded to the generalized BD condition preventing ghost gravitons in the branch corresponding to the AdS vacuum with cosmological constant $\Lambda_\star$. Thereby, unitarity of the CFT and the absence of ghosts gravitons in AdS, seem to be the two faces of the same holographic coin.

\subsection{Positivity of the energy and $3$-point functions}

The form of the $3$-point function of the stress-tensor in a CFT$_{d-1}$ is highly constrained. In \cite{OsbornPetkou,ErdmengerOsborn}, it was shown that it can always be written in the form
\begin{equation}
\langle T_{ab}(\mathbf x)\,T_{cd}(\mathbf y)\,T_{ef}(\mathbf z)\rangle=
\frac {\left( \mathcal A\, \mathcal I^{(1)}_{ab,cd,ef}+\mathcal B\, \mathcal I^{(2)}_{ab,cd,ef}+\mathcal C\, \mathcal I^{(3)}_{ab,cd,ef}\right)}{|\mathbf x - \mathbf y|^{d-1}\,|\mathbf y - \mathbf z|^{d-1}\,|\mathbf z - \mathbf x|^{d-1}} ~,
\label{3point}
\end{equation}
where the specific form of the tensor structures $\mathcal I^{(i)}_{ab,cd,ef}$ is irrelevant for us. Ward identities relate 2-point and 3-point correlation functions, which means that the central charge $C_T$ can be written in terms of the parameters $\mathcal A$, $\mathcal B$ and $\mathcal C$,
\begin{equation}
C_T=\frac{ \pi^{\frac{d-1}{2}}}{\Gamma\left[\frac{d-1}2\right]}\,\frac{(d-2)(d+1)\mathcal{A}-2\mathcal{B}-4 d \, \mathcal{C}}{(d-1)(d+1)} ~.
\end{equation}
Nicely enough, an holographic computation of the parameters entering the above formula can be tackled. It is certainly more intricate than that of $C_T$; thus we omit the details. The result is \cite{CamanhoEdelstein2}
\begin{equation}
\mathcal{A} = \frac{\Gamma[d]}{\pi^{d-1}} \left( a_1(d) \frac{\Upsilon'[\Lambda_\star]}{(- \Lambda_\star)^{d/2}} - a_2(d) \frac{\Upsilon''[\Lambda_\star]}{(- \Lambda_\star)^{d/2-1}} \right) ~,
\label{hologA}
\end{equation}
and analogous expressions for $\mathcal B$ and $\mathcal C$, where $a_i(d)$ are rational functions of $d$, the space-time dimensionality.

A convenient parametrization of the $3$-point function of the stress-energy tensor was introduced in \cite{HofmanM}. The idea is to consider a localized insertion of the form\footnote{Notice that we are splitting time and space indices and, thus, from now on vectors are understood as $(d-2)$ dimensional objects.} $\int d\omega\, e^{-i\omega t}\,\epsilon_{jk}\,T^{jk}(\mathbf x)$, and to measure the energy flux at light-like future infinity along a certain direction $\mathbf n$,
\begin{equation}
{\cal E}({\bf n}) = \lim_{r \to \infty} r^{d-2}\! \int_{-\infty}^\infty\!\!\! dt\; ~{\bf n}^i\, T^{\,0}_{\;~i}(t, r\, {\bf n}) ~.
\label{insertion}
\end{equation}
Given a state created by a {local gauge invariant operator} $\mathcal{O} = \epsilon_{ij}\, T_{ij}$, since $\epsilon_{ij}$ is a symmetric and traceless polarization tensor, the final answer for the energy flux is fully constrained by conformal symmetry to be \cite{CamanhoEdelstein1,BuchelEMPSS}
\begin{equation}
\langle \mathcal E(\mathbf n)\rangle = \frac{E}{\Omega_{d-3}} \left[ 1 + t_2 \left(\frac{|\mathbf n \cdot \mathbf\epsilon|^2}{|\epsilon|^2}-\frac 1{d-2}\right) + t_4 \left(\frac{|\mathbf n \cdot \mathbf\epsilon \cdot \mathbf n|^2}{|\epsilon|^2}-\frac 2{d(d-2)}\right) \right] ~,
\label{vevE}
\end{equation}
where $E$ is the total energy of the insertion, $\mathbf n \cdot \mathbf\epsilon = n_i\,\epsilon_{ik}$, $\mathbf n \cdot \mathbf\epsilon \cdot \mathbf n = n_i n_j\epsilon_{ij}$, and $|\epsilon|^2 = \epsilon^*_{ik}\epsilon_{ik}$, while $\Omega_{d-3}$ is the volume of a unit $(d-3)$-sphere. For any CFT$_{d-1}$, it is characterized by the two parameters $t_2$ and $t_4$. Being the quotient of 3-point and 2-point correlators, $\langle {\cal E}({\bf n}) \rangle$ is fully determined by the parameters $\mathcal A$, $\mathcal B$ and $\mathcal C$. In particular \cite{BuchelEMPSS},
\begin{equation}
t_2 = \frac{2d}{d-1}\frac{d(d-3)(d+1)\;\mathcal{A}+3(d-1)^2\;\mathcal{B}-4(d-1)(2d-1)\;\mathcal{C}}{(d-2)(d+1)\;\mathcal{A}-2\;\mathcal{B}-4d\;\mathcal{C}} ~,
\label{t2param}
\end{equation}
and a similar expression for $t_4$. We do not care about the latter for the following reason. If the CFT$_{d-1}$ is supersymmetric, $t_4$ vanishes \cite{HofmanM,KulaxiziP1}. On the other hand, even though there is no proof in the literature showing that Lovelock theories admit a supersymmetric extension, it turns out that the holographic computation suggests that a CFT$_{d-1}$ with a weakly curved gravitational dual whose dynamics is governed by Lovelock theory has a null value of $t_4$ \cite{CamanhoEdelstein2},
\begin{equation}
t_4 = 0 \qquad \Rightarrow \qquad \langle \mathcal E(\mathbf n)\rangle = \frac{E}{\Omega_{d-3}} \left[ 1 + t_2 \left(\frac{|\mathbf n \cdot \mathbf\epsilon|^2}{|\epsilon|^2}-\frac 1{d-2}\right) \right] ~.
\label{vevEfinal}
\end{equation}
The existence of a minus sign in (\ref{vevEfinal}) leads to interesting constraints on $t_2$, by demanding that the energy flux be positive for any direction $\mathbf n$ and polarization $\epsilon_{ij}$. For the tensor, vector and scalar channels, we obtain, respectively,
\begin{equation}
t_2 \leq d-2 ~, \qquad t_2 \geq - \frac{2 (d-2)}{d-4} ~, \qquad t_2 \geq - \frac{d-2}{d-4} ~. 
\label{t2channels}
\end{equation}
The vector channel constraint is irrelevant. In any supersymmetric CFT$_{d-1}$, therefore, the parameter $t_2$ has to take values within the window 
\begin{equation}
- \frac{d-2}{d-4} \leq t_2 \leq d-2 ~, 
\label{t2window}
\end{equation}
if the energy flux at infinity is constrained to be positive. For instance, any $\mathcal{N}=1$ supersymmetric CFT$_4$ has $|t_2| \leq 3$, with
\begin{equation}
t_2 = 6\,\frac{c-a}{c} \qquad \Rightarrow \qquad \frac{1}{2} \leq \frac{a}{c} \leq \frac{3}{2} ~.
\label{t2ac}
\end{equation}
where $a$ and $c$ are the parameters entering the trace anomaly formula, the bound being saturated for free theories \cite{HofmanM}.

\begin{figure}[h]
\includegraphics[scale=.7]{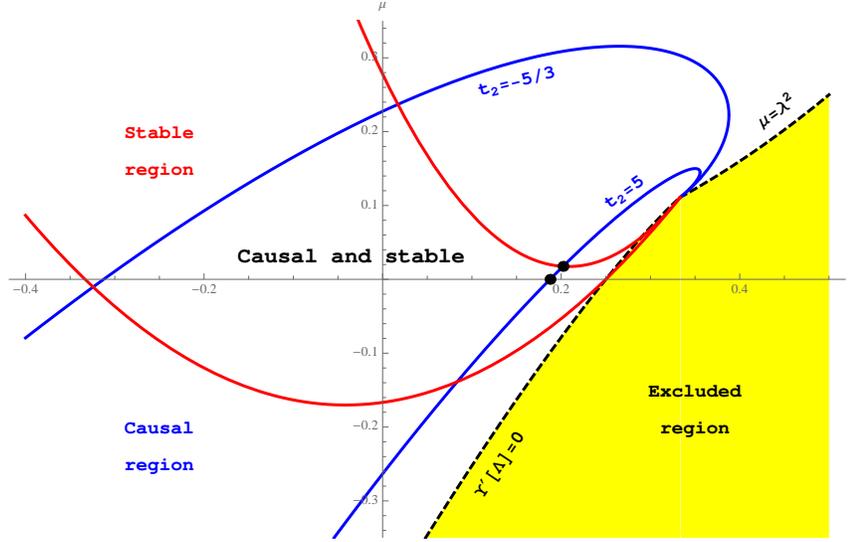}
\caption{The allowed region of gravitational couplings by causality and stability for cubic Lovelock theory in $d=7$ is displayed. The black points are the maximal values of $\lambda$ that can be attained in LGB \cite{deBoerKP1,CamanhoEdelstein1} and cubic Lovelock \cite{deBoerKP2,CamanhoEdelstein2} gravities, that are related to the lowest possible value of $\eta/s$ in a dual CFT$_6$ strongly coupled plasma \cite{CEP}.}
\label{LL3regions}
\end{figure}

We can holographically compute $t_2$ by inserting a shockwave on AdS, which sources the field theory insertion, and considering a metric fluctuation $h_{xy}(z,\mathbf x) := L_\star^2/z^2 \;\phi(z,\mathbf x)$ about this background. The $3$-point function follows from evaluating on-shell the effective action for the field $\phi$ on a particular shockwave solution. The relevant shockwave profile is   given by (\ref{sw2}), as discussed in \cite{HofmanM}. Up to an overall factor, the cubic vertex is \cite{BuchelEMPSS}
\begin{equation}
\mathcal{I}_{\rm cubic} \sim C_T \int\!d\mathbf x\,du\,dv ~\sqrt{-g} \,\phi\, \partial^2_v \phi\;\varpi\,\left( 1 - \frac{\Lambda_\star\,\Upsilon''(\Lambda_\star)}{\Upsilon'(\Lambda_\star)}\frac{T_2}{(d-3)(d-4)} \right) ~,
\label{cubicvertex}
\end{equation}
where 
\begin{equation}
T_2 = \frac{z^2 (\partial_x^2 \varpi + \partial_y^2 \varpi) - 2 z \partial_z \varpi - 4 \varpi}{\varpi} ~.
\label{T2}
\end{equation}
The relevant graviton profile \cite{BuchelEMPSS}
\begin{equation}
\phi(u=0,v,\mathbf x,z) \sim e^{-iEv}\,\delta(\mathbf x)\,\delta(z-1) ~,
\label{gravprofile}
\end{equation}
allows us to impose $\mathbf x = \mathbf 0$ and $z = 1$ in (\ref{T2}), this yielding the result
\begin{equation}
T_2 = 2 (d-1) (d-2) \left( \frac{n_x^2 + n_y^2}{2} - \frac1{d-2} \right) ~.
\label{T2final}
\end{equation}
We therefore read off, by plugging (\ref{T2final}) into (\ref{cubicvertex}) and comparing against the expression for $\langle \mathcal E(\mathbf n)\rangle$ in (\ref{vevEfinal}), the holographic prescription for $t_2$ in Lovelock theory:
\begin{equation}
t_2 = - \frac{2(d-1)(d-2)}{(d-3)(d-4)} \frac{\Lambda_\star\,\Upsilon''[\Lambda_\star]}{\Upsilon'[\Lambda_\star]} ~,
\label{t2holog}
\end{equation}
and $t_4 = 0$. Needless to say, this is the same expression we would have gotten by simply plugging the holographic formulas of the parameters $\mathcal A$, $\mathcal B$ and $\mathcal C$ (\ref{hologA}) into (\ref{t2param}). Combining (\ref{t2window}) and (\ref{t2holog}), we obtain \cite{deBoerKP2,CamanhoEdelstein2},
\begin{equation}
- \frac{d-2}{d-4} \leq -\frac{2(d-1)(d-2)}{(d-3)(d-4)} \frac{{\Lambda_\star}\,\Upsilon''[\Lambda_\star]}{\Upsilon'[\Lambda_\star]} \leq d-2 ~.
\label{LLwindow}
\end{equation}
For instance, Figure \ref{LL3regions} displays the two curves that establish the upper ($t_2 = 5$) and lower ($t_2 = - 5/3$) limits of the allowed window for the case of cubic Lovelock theory in $d = 7$. It enables us to appreciate how tight this restriction is in terms of the acceptable Lovelock couplings.
In the $d=5$ case, $-3 \leq t_2 \leq 3$, which together with the dependence of $t_2$ on the LGB coupling leads to \cite{BuchelM,Hofman}
\begin{equation}
t_2 = 12 \left( \frac{1}{\sqrt{1 - 4 \lambda}} - 1 \right) \qquad \Rightarrow \qquad - \frac{7}{36} \leq \lambda \leq \frac{9}{100} ~.
\label{t2vslambda}
\end{equation}
Notice, in particular, from (\ref{t2ac}), that $a = c$ corresponds to vanishing $t_2$ and $\lambda$. This suggests that higher curvature corrections are mandatory to study, for instance, four dimensional strongly coupled CFTs with $a \neq c$ under the light of the gauge/gravity correspondence.

\subsection{Gravitons thrown onto shock waves must age properly}

Consider a shock wave with profile given in (\ref{sw1}) in AdS with cosmological constant $\Lambda_\star$. We would like to analyze the following process. A highly energetic tensor graviton will be thrown from the boundary $z=0$ towards the shock wave.\footnote{The same computation can be carried out with vector and scalar gravitons, and the result in these two cases will be obvious from the present analysis.} This amounts to perturbing the relevant metric up to quadratic order and keeping only those terms involving derivatives like $\partial_u^2$, $\partial_v^2$ and $\partial_u \partial_v$ acting on the perturbation $\phi$:
\begin{equation}
\partial_u\partial_v\phi + \varpi_0\,L^2 \left( 1 + \frac{2(d-1)}{(d-3)(d-4)} \frac{\Lambda_\star\,\Upsilon''[\Lambda_\star]}{\Upsilon'[\Lambda_\star]} \right) \delta(u)\,z^{d-1}\,\partial_v^2\phi=0 ~,
\label{swpert}
\end{equation}
assuming $\Upsilon'[\Lambda_\star] \neq 0$, which means that the vacuum we are dealing with is non-degenerated. Causality problems arise when the coefficient of $\partial_v^2\phi$ becomes negative. In fact, notice that (\ref{swpert}) is a free wave equation except at the locus $u = 0$. We must only care about the discontinuity of $P_z$ for a graviton colliding the shock wave \cite{Hofman,CEP}
\begin{equation}
\Delta P_z = \frac{(d-1)}{z} |P_v|\, \left(\frac{z}{L}\right)^2 \, z^{d-3}\,\left(1 + \frac{2(d-1)}{(d-3)(d-4)} \frac{\Lambda_\star\,\Upsilon''[\Lambda_\star]}{\Upsilon'[\Lambda_\star]} \right) ~,
\label{DeltaPz}
\end{equation}
while the shift in the light-like time is \cite{CamanhoEdelstein2}
\begin{equation}
\Delta v =\left(\frac{z}{L}\right)^2\, z^{d-3}\,\left(1 + \frac{2(d-1)}{(d-3)(d-4)} \frac{\Lambda_\star\,\Upsilon''[\Lambda_\star]}{\Upsilon'[\Lambda_\star]} \right) ~.
\label{Deltav}
\end{equation}
Thus, if the quantity in parenthesis is negative, a graviton thrown into the bulk from the AdS boundary, bounces back, landing outside its own light-cone! This is understood as a signal of causality violation. If we repeat this computation for vector and scalar polarizations, we end up with the constant
\begin{equation}
- \frac{d-2}{d-4} \leq -\frac{2(d-1)(d-2)}{(d-3)(d-4)} \frac{{\Lambda_\star}\,\Upsilon''[\Lambda_\star]}{\Upsilon'[\Lambda_\star]} \leq d-2 ~.
\end{equation}
These are exactly the allowed values for $t_2$ --once the holographic dictionary has been put into work--, that ensure positivity of the energy in the dual CFT. This ends up, once again, in an alluring match between gravity and gauge theory.

\subsection{Black holes and plasma instabilities}

We could have obtained the results of the previous subsection following a different approach. Consider Lovelock black holes and study the potentials felt by high momentum gravitons exploring the bulk. Close to the boundary, $z \ll z_+$, for the different helicities \cite{deBoerKP2,CamanhoEdelstein2}
\begin{eqnarray}
c^2_{\rm tensor} & \approx & 1 + \frac{1}{L_\star^2 \Lambda_\star} \frac{z^{d-1}}{z^{d-1}_+} \left[1 + \frac{2(d-1)}{(d-3)(d-4)}\frac{\Lambda_\star\,\Upsilon''[\Lambda_\star]}{\Upsilon'[\Lambda_\star]}\right] ~, \\ [0.4em]
c^2_{\rm vector} & \approx & 1 + \frac{1}{L_\star^2 \Lambda_\star} \frac{z^{d-1}}{z^{d-1}_+} \left[1 - \frac{(d-1)}{(d-3)}\frac{\Lambda_\star\,\Upsilon''[\Lambda_\star]}{\Upsilon'[\Lambda_\star]}\right] ~, \\ [0.4em]
c^2_{\rm scalar} & \approx & 1 + \frac{1}{L_\star^2 \Lambda_\star} \frac{z^{d-1}}{z^{d-1}_+} \left[1 - \frac{2(d-1)}{(d-3)}\frac{\Lambda_\star\,\Upsilon''[\Lambda_\star]}{\Upsilon'[\Lambda_\star]}\right] ~.
\end{eqnarray}
The argument proceeds as follows \cite{BriganteLMSY2}. Notice that the potentials are normalized in such a way that their boundary value is $1$, while they need to vanish at the black hole horizon. These potentials can be understood as the square of the local speed of gravitons with the corresponding polarization. Even though there is no problem with a graviton whose local speed surpass that of light measured at the boundary, any excess would entail the existence of a local maximum.

Therefore, the graviton energy can be fine-tuned in such a way that it stays an arbitrarily large period of time at the top of the potential. Without the need of an explicit knowledge of the geodesic, it is clear that the average speed of the graviton will be bigger than the speed of light at the boundary. Since the graviton bounces back into the boundary, it means that there would be a corresponding excitation in the dual gauge theory that becomes superluminal. This should be forbidden in any sensible theory that respects the principle of relativity.

The conditions $c^2_{\rm tensor}$, $c^2_{\rm scalar} \leq 1$ in the vicinity of the boundary lead to the same constraints found before. We can argue that this is due to the fact that the shockwave analysis is related to the current one through a Penrose limit. A more physical interpretation would be that causality violation is not linked to the existence of a black hole solution since it is not due to thermal effects.

Once we consider the current setup, there is a second source for pathologies. If any of the squared potentials becomes negative anywhere, either close to the black hole horizon or deep into the bulk, an imaginary local speed of light will reflect an instability of the system. This ceases to exist in the absence of a black hole. Thus, it seems natural to identify it with a thermal feature of the CFT. They should correspond to plasma instabilities \cite{BuchelEMPSS}. Analogously to what happens with the restrictions coming from the window of allowed values for $t_2$, these restrictions further constrain the values that Lovelock couplings can take in a sensible theory \cite{CEP}. This is explicitly shown in Figure \ref{LL3regions} for the case of cubic Lovelock theory in $d = 7$. There, the region of Lovelock couplings leading to causal and stable physics is given by a connected and compact vicinity of the EH-point ($\lambda = \mu = 0$).

\section{Final comments}

The study of higher curvature gravity in the context of the AdS/CFT correspondence appears, at the least, as a territory worth exploring. It allows to further understand how profound concepts of quantum field theory might be linked, holographically, to comparable deep concepts in the realm of gravity. Some examples were briefly presented above, such as the relation between positivity of the energy in the CFT and a certain kind of causality violation in the dual gravitational theory. We have not discussed, although they exist \cite{CEP}, other sources of causality violation occurring in the bulk, which do not seem to be related to pathologies inherited from the $3$-point stress-energy tensor correlators in the dual CFT.

Lovelock theories are remarkable in that lots of physically relevant information is encoded in the polynomial $\Upsilon[g]$. BD instabilities, for instance, can be simply written as $\Upsilon'[\Lambda_\star] < 0$, which has a beautiful counterpart telling us that the central charge of the dual CFT, $C_T$, has to be positive. This is unitarity. Now, $\Upsilon'[\Lambda_\star]$ is the asymptotic value of the quantity $\Upsilon'[g]$, the latter being meaningful in the interior of the geometry, and positive along the corresponding branch. Recalling that naked singularities take place at extremal points of $\Upsilon[g]$ further suggests that $\Upsilon'[g]$ might be a meaningful entry in the holographic dictionary (see \cite{Paulos} for related ideas).

In spite of the higher dimensional nature of Lovelock theory, it is important to mention that there are lower dimensional gravities, dubbed {\it quasi-topological}, whose black hole solutions are alike those discussed in this talk. Many of the results presented above are pertinent in those ``more physical'' setups of AdS/CFT \cite{MyersPS}. 

\begin{acknowledgement}
I am very pleased to thank Xi\'an Camanho, Gast\'on Giribet, Andy Gomberoff and Miguel Paulos for collaboration on this subject and most interesting discussions held throughout the last few years. I would also like to thank the organizers of the Spanish Relativity Meeting in Portugal ({\sc ERE2012}) for the invitation to present my work and for the nice scientific and friendly atmosphere that prevailed during my stay in Guimar\~aes. This work is supported in part by MICINN and FEDER (grant FPA2011-22594), by Xunta de Galicia (Conseller\'{\i}a de Educaci\'on and grant PGIDIT10PXIB206075PR), and by the Spanish Consolider-Ingenio 2010 Programme CPAN (CSD2007-00042). The Centro de Estudios Cient\'\i ficos (CECs) is funded by the Chilean Government through the Centers of Excellence Base Financing Program of Conicyt.
\end{acknowledgement}

\end{document}